# Blockchain-based Optimized Client Selection and Privacy Preserved Framework for Federated Learning

Attia Qammar[1] · Abdenacer Naouri[1]. Jianguo Ding[2] · Huansheng Ning[1]

[1] School of Computer and Communication Engineering, University of Science and Technology Beijing, Beijing, China
Email: B20200693@xs.ustb.edu.cn / q.attia@yahoo.com , nacer.naouri@gmail.com ,
ninghuansheng@ustb.edu.cn

[2] Department of Computer Science, Blekinge Institute of Technology, Karlskrona, Sweden
Email: jianguo.ding@bth.se
Corresponding author: Huansheng Ning

**Abstract:**
Federated learning is a distributed mechanism that trained large-scale neural network models with the participation of multiple clients and data remains on their devices, only sharing the local model updates. With this feature, federated learning is considered a secure solution for data privacy issues. However, the typical FL structure relies on the client-server, which leads to the single-point-of-failure (SPoF) attack, and the random selection of clients for model training compromised the model accuracy. Furthermore, adversaries try for inference attacks i.e., attack on privacy leads to gradient leakage attacks. We proposed the blockchain-based optimized client selection and privacy-preserved framework in this context. We designed the three kinds of smart contracts such as 1) registration of clients 2) forward bidding to select optimized clients for FL model training 3) payment settlement and reward smart contracts. Moreover, fully homomorphic encryption with Cheon, Kim, Kim, and Song (CKKS) method is implemented before transmitting the local model updates to the server. Finally, we evaluated our proposed method on the benchmark dataset and compared it with state-of-the-art studies. Consequently, we achieved a higher accuracy rate and privacy-preserved FL framework with decentralized nature.
**Keywords:** Federated learning, Blockchain, Smart contracts, Privacy, Optimized client selection

## 1. Introduction

Federated learning (FL) is a kind of distributed machine learning (ML) paradigm that ensures data privacy without centralized data storage. In traditional FL all data are trained locally and sent to the server as the trained local model updates and then local model updates are aggregated into the global FL model [1]. Since it is not required to upload the user's raw data in order to prevent data privacy leakage during transmission and model training. In recent years, FL has become a hotspot research area in academia and industry due to its privacy-preserved nature. Although FL prevents the raw data leakage of contributed FL participants, however, it causes a single-point-of-failure (SPoF) attack. Besides, studies presented the federated learning attack surface that comprises various attacks including, poisoning attacks, inference attacks, free-riding attacks, and many others [2]–[5]. Furthermore, in a conventional FL environment, the participants are randomly selected for the model training. The randomly selected participants may have heterogeneous data resources and computational power. Due to these reasons, FL participants may lead to drop out during FL model updates transmission. Additionally, participants can send malicious model updates that cause data poisoning attacks as well as privacy leakage issues. In order to select the particular participants for FL model training is required to register with the unique identifiers and impose penalties in case of notorious activity.

Recently, a couple of studies are investigated to provide optimum client selection and to motivate them with incentives in conventional FL [6]–[8]. However, these studies focused on the centralized server for model transmission and storage which possibly leads to the SPoF attack, biased and malicious model. Besides, randomly selected FL clients can intentionally jeopardize the FL environment by threatening the global FL model [9]. Furthermore, attackers exploited the gradient information of the FL model to obtain sensitive information. In order to cope with these limitations in the FL ecosystem we required a blockchain-based optimized client selection and privacy-preserved federated learning system. The integration of Blockchain technology into federated learning has the capability to solve the above problems in decentralized nature and maintain the record [10]–[14]. Furthermore, homomorphic encryption-based techniques are employed to protect the model from malicious users [15], [16]. In the homomorphic encryption method, model updates are encrypted and FL aggregation operation is performed on the encrypted updates. In the work of [17], authors implemented the ElGamal, in which message expansion by the factor of two takes place during encryption, which means the ciphertext is twice as long as the plaintext. Furthermore, it is a partially homomorphic scheme, which only allows multiplicative homomorphic encryption. Whereas in the FL process, additive homomorphic encryption is required to aggregate the local model updates, although it is possible to slightly modify ElGamal's multiplicative homomorphic encryption into the additive. However, it is not possible to perform both operations at the same time [18], [19]. Hence, we decided to implement the CKKS FHE scheme, which can provide addition and multiplication operations at the same time on large-scale datasets. Conclusively, the main contributions of our paper are as follows:

- In order to select the optimized clients for FL training a forward bidding auction method is presented based on the smart contract (SC), registered the FL clients with unique addresses, placed the bid according to the requirements announced by the FL task publisher/server and deposit some amount as security by registered FL clients.
- To alleviate the data protection, we applied the fully homomorphic encryption on model weights before sending it to the FL server and the aggregation operation is performed on the encrypted data. Furthermore, the presented framework uses the Ethereum blockchain and hash of averaged global FL model stores on the IPFS to provide decentralization storage.
- Finally, the performance of our proposed framework is evaluated through simulation on benchmarked open real-world datasets.

The remainder of the paper is organized as follows: Section 2, presented the related work. Section 3, describes the system model and problem statement. In section 4, we elaborated the proposed blockchain-based optimized client selection and privacy-preserved framework for federated learning. Section 5, investigates the results. Finally, section 6 provides the conclusion and future direction.

## 2. Related Work

Recently, federated learning and blockchain research have been developed immensely, and privacy leakage during data transmission has also increased critically. Federated learning provides data protection across data sharing, as authors [20] presented the FL framework to enable privacy in shared models with distributed Internet of Things (IoT) networks. Fu et al. [21] utilized the FL in privacy-preserving manners for industrial big data processing and the Lagrange interpolation method is implemented to verify the accuracy of aggregated gradients. Additionally, authors in [22] proposed the edge-cloud assisted FL model to communicate efficiently with secure energy data sharing and Song et.al [23], optimized the privacy of users' identity in the FL environment for mobile edge computing. However, in the federated learning process, information leakage can be occurred by malicious clients or servers which leads to inference attacks [24][25]. Additionally, the traditional FL structure depends on a centralized server and is vulnerable to a single-point-of-failure attack [10], [26].

In this context, blockchain technology has the potential to integrate with federated learning and well manage the aforementioned problems. The blockchain-based FL federated learning can improve the model efficiency and solve data island issues. Qammar et al. [12], presented a comprehensive systematic review to explain the integration of blockchain with federated learning in order to provide a secure FL environment. Kim et al. [27], presented the blockchain-enabled federated learning as BlockFL to coordinate the model learning process in a decentralized manner. Besides, the authors investigated the end-to-end model latency of BlockFL and classify the optimized block generation rate by taking into account communication, computation, and consensus delays. Similarly, authors [28], [29] implemented the smart contract to enable the reputation management of FL clients and maintain the reputation history through the blockchain ledger, hence comprehending the tamper resistance nature of the reputation. Furthermore, Qu et al.[30], integrated the blockchain with federated learning aim to select fair nodes to participate in particular FL model training tasks in Industry 4.0.

Zhang et al. [17] proposed the masking methods based on homomorphic encryption (HE) and the secure multi-party computation (SMC) for FL in order to securely calculate the model weights by considering the data quality of updates in comparison with the conventional FL weight calculation method. Additionally, the authors implemented the ElGamal algorithm to encrypt local model updates. In the work of authors [31], proposed the blockchain-enabled FL for the Industrial Internet of Things (IIoT) to formulate secure data aggregation in terms of model sharing. Authors applied multiple security including differential privacy (DP) and homomorphic encryption (HE) with distributed K-mean clustering i.e., random forest and AdaBoost, respectively in a byzantine IIoT environment. Moreover, in the studies of TrustFed [32], researchers exploited the blockchain smart contract in federated learning to maintain the reputation of client devices in a way of honest and active model contributions. Besides, TrustFed efficiently detects and removes outliers and guarantees fair FL model training.

A Federated learning system required a trustworthy environment to successfully implement the AI principle. These studies have concentrated on the communication, computation efficiency, reputation, incentives, and data heterogeneity issues of federated learning. However, few researchers have systematically focused on the security and privacy of federated learning based on the blockchain in a decentralized manner. Based on the previous studies, this paper works towards the blockchain-based optimized client selection and privacy-preserved FL framework by introducing the forward auction method, registering FL clients by a smart contract, applying the aggregation process on encrypted model updates, and storing the FL model in a decentralized environment.

## 3. System Model and Problem Statement

Federated learning trains the millions of artificial intelligence models that learn from jointly shared models with the help of participating devices. FL ensures the model training with a couple of FL participants or clients without distributing the raw training data. Generally, federated learning works in three steps such as 1) initial model requirements are sent to the all-participated clients in the federated learning system, 2) local model updates are sent to the server by participated clients, and 3) aggregated global model is produced by the server with the aggregation of all local model updates as presented in Figure 1. The total number of clients participating in the FL are supposed to be $N$ and $n_i$ belongs to the number of samples from the individual client. At time $t$, FL server selects few clients such as $k.N$ ($0 < k \leq 1$) randomly from all participated clients and then further asks them to update the global model with the local dataset. According to authors [1], federated learning is mathematically presented in equation (1) with the Federated Averaging (FegAvg) algorithm.

$$w_{t+1} = w_t + \eta \cdot \frac{\sum_{i=1}^{kN} n_i \Delta w_{t+1}^i}{\sum_{i=1}^{kN} n_i} \quad (1)$$

In above Eq. (1) $w_t$ presents the previous model updates at time $t$, whereas $w_{t+1}$ is the updated model at time $t + 1$. Furthermore, $\Delta w_{t+1}^i$ shows the updated change in global model parameters by client $i$. At last, $\eta$ is the learning rate of global FL model.

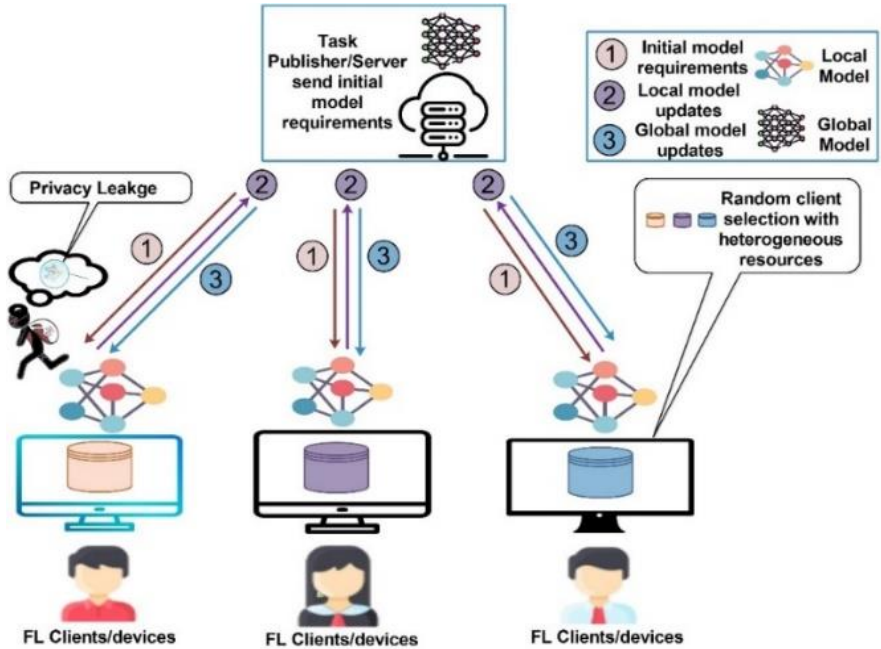

**Figure 1: Federated learning process with inherited problems**

The main problem is that the clients are randomly selected for model training, and at every stage during model uploading can be attacked by malicious users in the FL process [1]–[4]. Similarly, a random selection of clients leads to an imbalanced data distribution which negatively affects the model performance. Moreover, it makes the FL system more vulnerable to unauthorized access by malicious clients that causes inference attacks leading to potential data leakage and privacy violations. The randomly selected clients may have low computational resources, less bandwidth, and leaves the training process in the middle.

Several federations have stringent privacy laws or regulations that may require the system to include extra security measures to secure against the inference of personal data. Under some circumstances, sending unencrypted updates to the model could expose confidential information to potential adversaries. In federated learning, data kept to be local is not enough because attackers used the gradient information to extract sensitive information. Hence, we consider these problems in our research and proposed the blockchain-based optimized client selection with privacy preserved federated learning framework.

## 4. Proposed Method

In this section, we proposed the blockchain-based optimized client selection and privacy-preserved framework for federated learning. The proposed system consists of the following components i.e., task publisher/server, FL clients/devices, integration layer, and blockchain layer as presented in Figure 2.

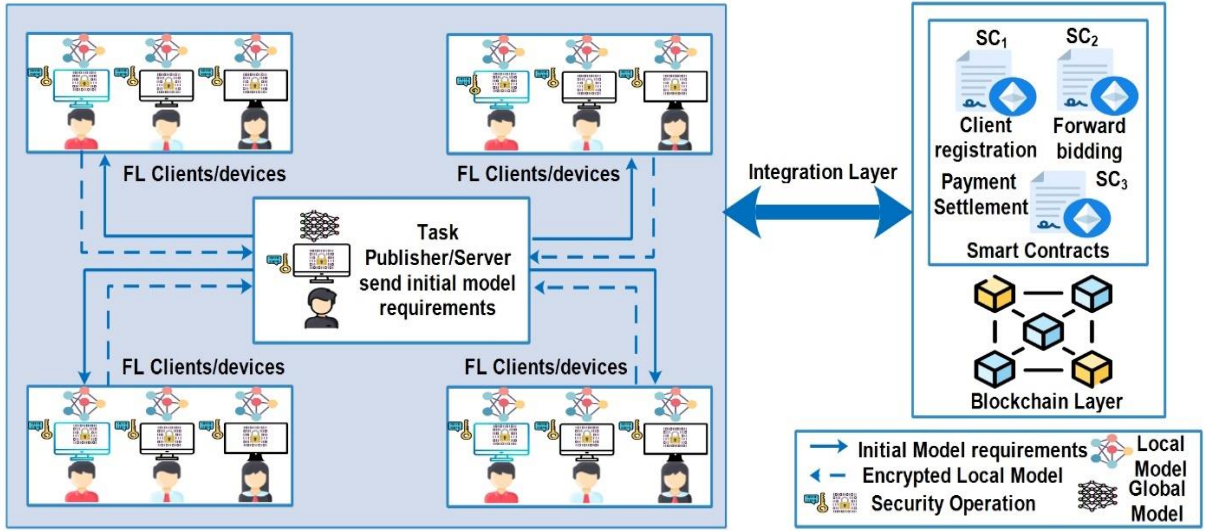

**Figure 2: Blockchain-based optimized client selection and privacy-preserved framework**

1. **Task Publisher/Server:** The FL task publisher or server have a responsibility to announce the task and requirements to complete it by opening the auction process. After that, it sends the initial model parameters to the whitelisted FL clients/devices who are already registered based on the predefined criteria, aggregates the all-encrypted local model updates into the global model, saved the model at decentralized storage i.e., IPFS, and sends the hash of the global model to the blockchain network.
2. **FL Clients/Devices:** FL clients/devices send the request for registration, once they are registered with unique addresses then they can join the auction to place the bid. After successfully placing the bid, the FL client/devices are selected for model training.
3. **Integration Layer:** The integration layer enables the connection between blockchain and decentralized applications (DApps). For instance, Web3.py originally derived from Web3.js is used to interact with the Ethereum blockchain network [33]. It usually helps to send transactions, read data from blocks, interact with smart contracts, and many other use cases.
4. **Blockchain layer:** The blockchain layer enables the different smart contracts (SC) functionality and storage of the model. We employed the Ethereum blockchain network and the smart contracts are written in solidity. In our paper, three kinds of smart contracts are developed such as $SC_1$, $SC_2$, and $SC_3$ for FL client registration, forward bidding, and payment settlement with reward, respectively.

The registration SC allows to register of FL clients with unique addresses in the system. These registered FL clients start placing the forward bidding when an auction is open. Based on the predefined criteria only already registered FL clients/devices can place the bid. After the placement of bids and the selection of Top X FL clients/devices, model training is started with encrypted model updates. When FL model training is finished and achieved the fully converged global model, then the payment settlement SC function is activated to return the deposited security amount to FL clients\devices and the remaining service amount to the task publisher\server. Moreover, we utilized another third-party storage such as the interplanetary file system (IPFS) [34] to store the FL model in a decentralized manner because blockchain is not suitable for maximum data storage. IPFS is a content-addressable and peer-to-peer system that stores the data and generates a hash as the address. The hash of the global FL model is saved on the blockchain node to maintain the record. In Figure 3, the workflow of the entire process is presented.

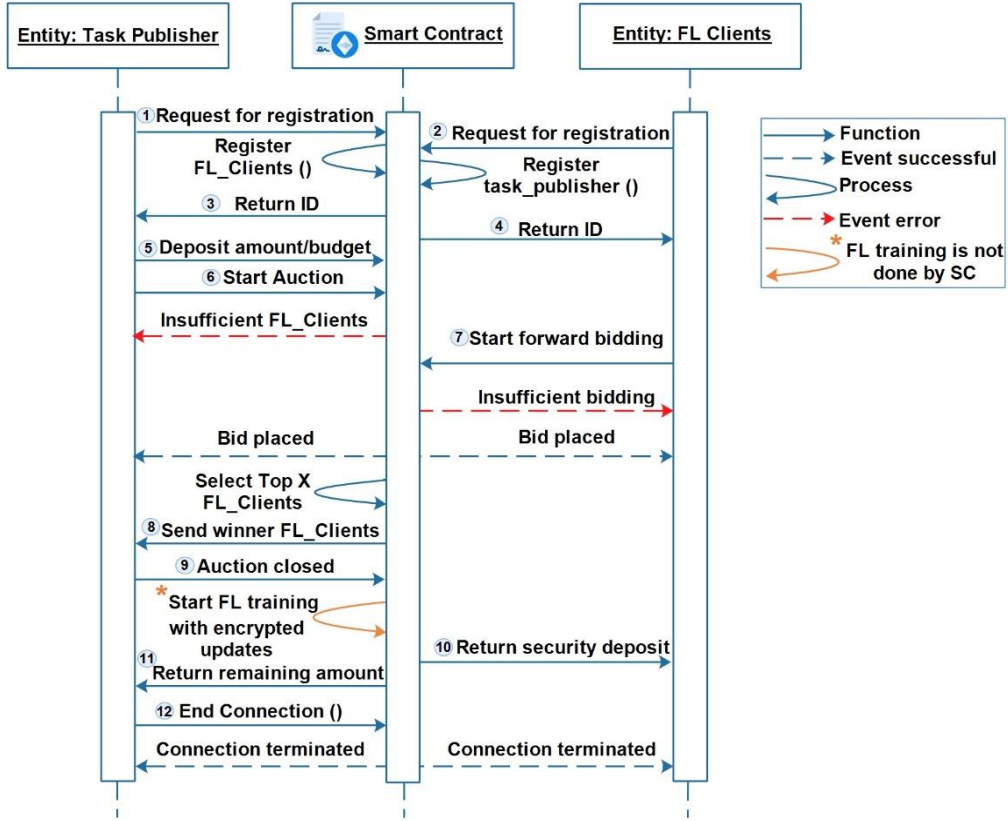

**Figure 3:** Sequence diagram of the blockchain-based optimal client selection

In Table 1, we have presented the list of notations with description used in our paper.

**Table 1: List of Notations**

| Notations | Description |
|---|---|
| $w_t$ | Presents the previous model updates at time $t$ |
| $w_{t+1}$ | Updated model at time $t+1$. |
| $\Delta w_{t+1}^i$ | Updated change in global model parameters by client $i$. |
| $\eta$ | Learning rate of global FL model |
| $\wedge$ | And Operator |
| $>=$ | Greater than or equal to |
| $<=$ | Less than or equal to |
| $FL_c$, | Federated learning client |
| $FL_s$, | Federated learning server |
| $PK$ | Public key |
| $SK$, | Private key |
| $(W_0)$ | Global model parameter |
| $E(W_0, PK_{FL_c})$ | Initial encrypted global model |
| $E(W_{n^{FL_c}}, PK_{FL_c})$ | Encrypted model by FL client $FL_c$ after "n" iterations |
| | Aggregation function agg() |
| $\Delta E(+ W_{n^{FL_c}} \ldots \ldots, PK_{FL_c})$ | Encrypted aggregated global model |
| $E(W_{n+1}) \; D(E(W_{n+1}, PK_{FL_c}), SK_{FL_s})$ | Decrypted global model at server side |

## 4.1 Forward Bidding Process

We have presented a forward-bidding smart contract auction method in order to select Top X-optimized federated learning clients that participated in the FL model training. The task publisher/server announced the auction over the Ethereum blockchain network for FL clients where they send bids to the contract. Additionally, the task publisher communicates the budget, security deposit, required computational resources, bandwidth, dataset size, dataset type, number of iterations and etc. to train the federated learning global model as elaborated in the algorithm one from steps 1 to 4. Furthermore, the deposited security amount is collected as a guarantee so that FL clients cannot leave or drop out of the training process otherwise, they will lose their amount. The task publisher returned the deposited amount to FL clients once the model training is finished.

Before sending the bids to the smart contract, FL clients are registered with a unique address as mapped to the Ethereum Address (EA). Only registered FL clients/devices start placing the forward bidding with the aforementioned requirements announced by the task publisher within the described time duration. The registered clients are put into a whitelisted/winners' section and chosen for the next round of iterations with predefined requirements. When the required number of bids have been placed as well broadcast the new bid to all clients, update the forward bidding record to all clients and then finally task publisher only selected the top X FL devices as the winner to participate in the FL model training as presented in algorithm one from steps 5 to 14. After that, reduced the amount equal to all offered bid prices from the FL task publishers/server account and closed the auction as elaborated in algorithm one in steps 15 and 16, respectively.

**Algorithm 1: Forward Bidding Process**

**Input**: task_publisher address, all requirements to participate in FL training for OnlyFL_clients
**Output:** TopX registered FL_clients are selected through a forward bidding process

1. task_publisher address is an Ethereum Address (EA)
2. *Modifier*: OnlyFL_clients who want to participate in FL_training
3. **If** FL_clients are registered $\wedge$ forward_bidding is open for bidding $\wedge$ Ether transferred = rate $\wedge$ task_publisher computational_resources, bandwidth, datatype, datasize, number of iterations $>=$ computational_resources, bandwidth, datatype, datasize, number of iterations **then**
4.    **If** a previous bid has been placed, **then**
5.      **If** offered resources $<=$ previous bid **then**
6.        revert
7.      **end**
8.    **else**
9.      **If** offered resources $>$ previous bid **then**
       Increase the total number of FL_clients
10.      Continue forward bidding of current FL_clients
11.      Broadcast the new bid to all participating FL_ clients
12.      Store all offered bids
13.      Update the forward bidding data to the new FL_ client
14.      Select top X FL_clients bids **then**
15.      Reduce task_publisher amount equals to all offered bid price
16.      Auction closed
17. **else**
18.    revert
19. **end**

## 4.2 FL Training Process with CKKS Encryption Scheme

After the selection of top X FL clients, the model training is started in a privacy-preserving manner. The model training is done by each client end and they send the encrypted local model updates to the server. We applied the Cheon-Kim-Kim-Song (CKKS) fully homomorphic encryption (FHE) [35] scheme on model weights before sending them to the FL server. In the work of [17], authors implemented the ElGamal, in which message expansion by the factor of two takes place during encryption, which means the ciphertext is twice as long as the plaintext. Furthermore, it is a partially homomorphic scheme, which only allows multiplicative homomorphic encryption. Whereas in the federated learning scenario, we also need the additive homomorphic encryption to aggregate the local model updates, although it is possible to slightly modify ElGamal's multiplicative homomorphic encryption into the additive. However, it is not possible to perform both operations at the same time [18], [19]. Hence, we decided to implement the CKKS FHE scheme, which can provide addition and multiplication operations at the same time on large-scale datasets.

The FL clients employed the public key to encrypt the model updates, after aggregation, the task publisher\server sends the aggregated model back to FL clients, and the process remains to continue until achieved the fully converged global model as presented in algorithm 2 from steps 1 to 5. At the client end, the local model builds by using Convolutional Neural Network (CNN) model with the real-world dataset. Moreover, in case of abandoning the FL training process, FL_clients lose their security deposit as a penalty otherwise they get the reward in the form of a fully converged global model as elaborated in algorithm 2 from steps 6 to 12.

**Algorithm 2: FL Training Process with CKKS Fully Homomorphic Encryption (FHE) Scheme**

**Input:** Task_publisher/FL server sends the initial encrypted model details to the selected Top X FL_clients

**Output:** FL_server has received the encrypted model updates aggregate them and produced the fully converged global model

Suppose that, the federated learning client $FL_c$, federated learning server $FL_s$, the homomorphic encryption keys public and private keys $PK$ and $SK$, respectively, and initial global model parameter $(W_0)$

1. **If** federated learning server $FL_s$ sends the initial encrypted global model $E(W_0, PK_{FL_c})$ to the Top X federated learning clients $FL_c$ **then**
2. Each client $FL_c$ trains the initial encrypted model $E(W_0)$ based on their local dataset
3. and sends back to the central server $FL_s$ as encrypted model $E(W_{n^{FL_c}}, PK_{FL_c})$ by FL client $FL_c$ after "n" iterations
4. The federated learning server $FL_s$, applied the aggregation function agg () on all encrypted local model updates as $\Delta E(W_{n^{FL_c}} + W_{n^{FL_c}} \dots \dots, PK_{FL_c})$ and produced the aggregated global model $E(W_{n+1})$ and $D(E(W_{n+1}, PK_{FL_c}), SK_{FL_s})$ decrypted global model at server side.
5. Repeat steps 1 to 3 until fully converged global model produced
6. **If** FL_clients left the training process **then**
7. Lose their deposit as penalty
8. **else**
9. Received the fully converged global model
10. **else**
11. revert
12. **end**

## 4.3 FL Model Training Completed and Payment Settlement

In algorithm 3, forward bidding is closed and the payment settlement method is activated. Upon successful FL model convergence, service amount is transferred to the FL clients and the hash of the global model shared with the FL clients as the reward. Finally, closed the bidding status, the service amount is reduced from task publisher\server account, broadcast the closing status, refund the remaining deposited amount to task publisher and the security fee to the registered participated FL clients. However, connection cannot be terminated before the FL model training and payment clearance as described in steps 2 to 13 from algorithm 3.

**Algorithm 3: FL model training completed and payment settlement SC**

1. **Modifier:** task_publisher
2. **If** ∧ Closing time of forward bidding has been passed ∧ sufficient bidding offers has been made ∧ FL model training completed **then**
3. Transfer the service fee to the FL_Clients $FL_c$,
4. Shared the hash of fully converged global as reward with FL_Clients
5. Set forward bidding status is closed

| | |
|---|---|
| 6. | Decrease task_publisher deposit according to service charges |
| 7. | Broadcast the forward bidding status is closed |
| 8. | Refund the remaining deposit amount to the task_publisher |
| 9. | Transfer the security deposit to registered participated FL_Clients |
| 10. | Terminate the connection |

11. **else**
12.    revert
13. **end**

In FL-MAB [20], authors returned the deposit to FL clients after the bidding process was completed and then started the FL training round, it is possible some FL clients drop out during model training, there doesn't seem to be a penalty mechanism to discourage clients from dropping out. Additionally, it does not provide any explicit privacy guarantees for the local model updates transferred to the FL server. Similarly, rewards for the FL clients based on their contribution to the global model that may not always be fair or reasonable. Because FL clients with large datasets and high computational resources may have an unfair advantage, leads to an unequal distribution of rewards. However, we proposed the forward bidding mechanism to select top X FL clients within the budget of FL task publisher\server and paid them accordingly. In order to ensure the FL clients do not drop out during model training, they submitted the fixed security amount and returned them after successfully converged the global model with reward. Furthermore, to guarantee the privacy of FL model we applied the fully homomorphic encryption scheme with CKKS. In this way FL operations are applied on the encrypted data and ensures the privacy.

## 5. Results and Analysis

### 5.1 Setup and datasets

We performed the experiment on the system equipped with (an Intel®CoreTM i7-8650U @ 2.10 GHz CPU, and 32 GB RAM running on Windows Server 2019). The federated learning environment is implemented using the Python 3.9. with PyTorch framework. We performed the federated learning model training with Convolutional Neural Network (CNN) model, the number of clients considered the 30 with forward bidding, the number of epochs is between 10 to 50, and stochastic gradient descent (SGD) optimizer used with momentum to improve the performance of neural network. The other parameters such as learning rate and local batch size considered the 0.01 and 10, respectively. Furthermore, smart contracts are written in solidity programming language and Truffle framework with Ganache used for deployment and testing. The Mythril is used to audit and validation of smart contract in terms of security analysis. To implement the fully homomorphic encryption, we utilized the Pyfhel library with CKKS scheme. Besides, web3. Py provides the interaction with Ethereum blockchain and global model stored on the IPFS which is decentralized storage with Infura API.

In this work, we performed experiment with MNIST dataset with Independent and identically distributed (IID) strategy. The MNIST dataset contains 60,000 greyscale images with 28*28 pixels,10 classes, and have 10,000 test images and 50,000 training images of digits dataset from 0 to 9.

### 5.2 Performance Evaluation Metrics

To evaluate the performance of the proposed method, we consider the following metrics accuracy, computation time, cost of operations, and compare with other state-of-the-art studies. The high value of the classification accuracy means that proposed model works well. Moreover, the computational time is analyzed for encrypted model updates and unencrypted model updates. The higher the computational time indicates that model updates are encrypted and it take more time to apply the FHE operation on it as compare to unencrypted model. The gas cost of the operations is measured for Ethereum blockchain at average and minimum rate.

### 5.2.1 Accuracy

In this experiment we first evaluate the classification accuracy of our proposed scheme, where we have compared the accuracy of optimized clients selected by forward bidding smart contract and randomly selected clients for federated learning model training. In Figure 4, it is presented that FL model training with optimized client selection achieved the higher accuracy of 98.36% as compare to randomly selected clients with MNIST dataset. With respect to different communications rounds i.e., 10, 20, 30, 40, and 50 the accuracy is increased with static fraction of 30 clients, however due to client's dropout, low computational resources, and less bandwidth etc., in random selection decreased the accuracy and only achieved the 97.53% till 50 communication rounds and continuously fluctuating and obtained the lower accuracy 87.29%. Whereas, in optimal client's selection maintain the accuracy after the $10^{th}$ round of communication. In our approach, optimized selection of clients improved the accuracy of model from 88.91% to 98.36% as increasing the communication rounds from 10 to 50.

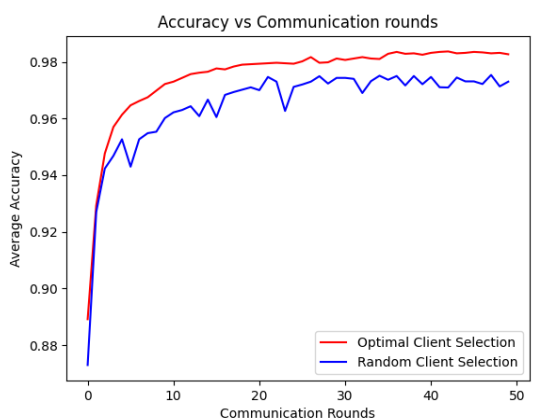
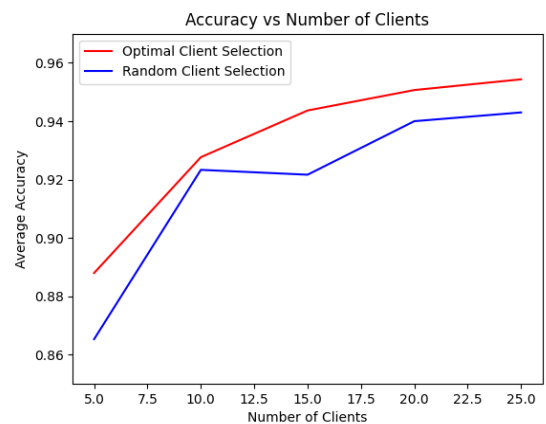

**Figure 4. Accuracy as the increase of communication rounds for optimal vs random selection of clients**

**Figure 5. Accuracy as increase the density of clients for optimal vs random selection of clients**

In Figure 5, we tested the accuracy of model by increasing the density of clients i.e., number of clients considered as 5%, 10%, 15%, 20% and 25% etc. We noticed that model accuracy increased with more density of clients, when we have 20% clients then accuracy is 94.30%, and maximum accuracy achieved 96.21% with the 25 number of clients. However, in random selection of clients still accuracy is less as compared to optimal selection of clients due to less computational resources, low bandwidth, less data size, etc. In random selection with 20% and 25% of clients obtained the 94.30% and 96.21% accuracy, respectively, which is significantly low as compare to optimal selection of clients.

Consequently, we proved that our proposed methods outperform in comparison with random selection of clients and other state-of-the-art baseline studies by selecting optimized resources in a decentralized manner.

### 5.2.2 Computation Time

We evaluated the average execution time of the selected optimal clients and random clients during the training process, taking into account the data size. We considered two scenarios: one with the same data model size and another with different sizes. Optimal clients were chosen based on their sufficient computing resources. We observed that the average execution time of the optimal clients was lower compared to the random clients for all communication rounds (2, 4, 6, 8, and 10), with different data size as shown in Figure 6. However, when training the model with random clients, it took more execution time due to some clients dropping out in the middle of training and lacking sufficient computing power to train the model efficiently. We consider the data size between i.e., 3000 - 4000 KB, the maximum completion time for optimal clients takes 14 sec, whereas random clients take the 38 sec, which is quite high.

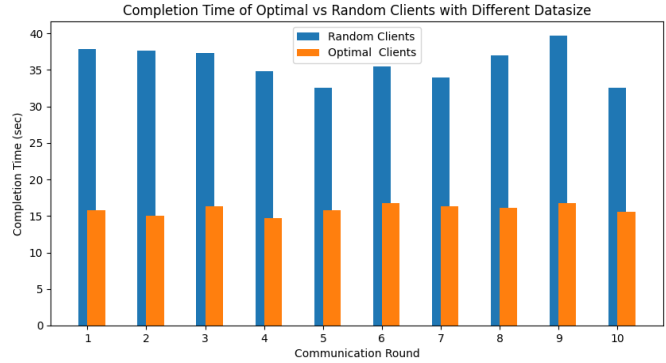

**Figure 6: Completion time as increasing the communication rounds with different data size in optimal vs random clients**

Furthermore, in Figure 7, we depicted the execution time of local model training with similar data size for both optimal and random client selection. Consistently, the completion average time of the optimal clients was always lower than the random clients. We considered the same data size as 3220 KB with the communication rounds 1 to 10. Optimal clients take only 12 (sec) to complete the local model training at their own end. For random clients, it takes more execution time around 29 (sec) and at round 4 execution time is at peak due to less computing power as well as some clients leaved the training process. Conclusively, optimal clients have better resources which are selected by smart contract with forward bidding process completed the local model training in minimum time.

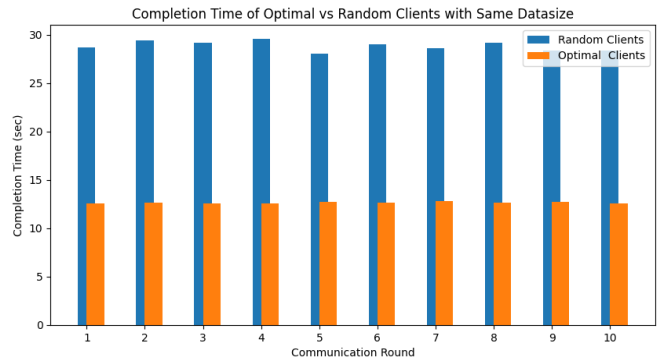

**Figure 7: Execution time as increasing the communication round with same data size in optimal vs random clients**

### 5.2.3 Execution Time and Accuracy with CKKS Fully Homomorphic Encryption

We applied the fully homomorphic encryption with CKKS on the local model updates to encrypt them and prevent from gradient leakage attacks. The CKKS conforms various parameters to ensure the high security level and performance. However, it is a trade-off between security and efficiency.

Similarly, CKKS involves the polynomial degrees which allows precise computations however, utilized the high computational resources and effects the performance. According to our requirements we set the polynomial degree to 8192 and it takes maximum execution time 4968 (sec) with 50 communication rounds as presented in Figure 8. As increasing the communication rounds it takes more time due to the implementation complexity of the fully homomorphic encryption to process encrypted data. It is the significant difference between the execution time with CKKS encryption and without encryption. However, without CKKS FHE with the same number of communication rounds, execution time is 3168 (sec), which is significantly less. Similarly, with 30 communication rounds with CKKS takes 2893 (sec) and without CKKS takes 1493 (sec), which almost half of the CKKS encryption time.

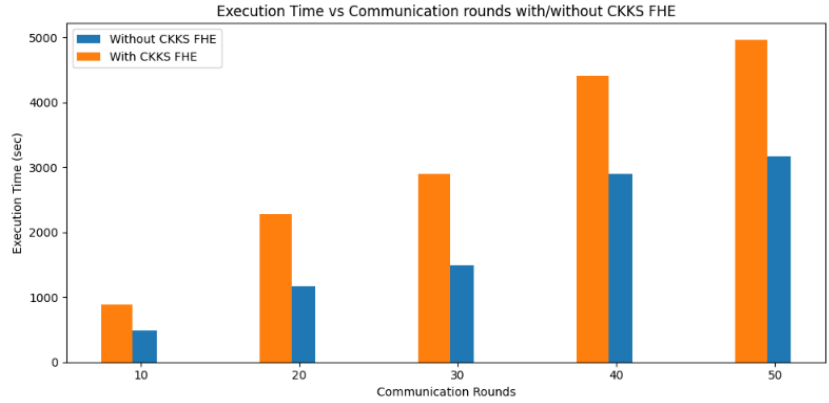

**Figure 8: Execution time with and without CKKS FHE**

Furthermore, in Figure 9, we presented the accuracy it decreased with CKKS encryption, and achieved the maximum accuracy 97.23% with 50 communication rounds. However, without encryption we obtained the 98.36% accuracy with 50 communication rounds. Similarly, lowest accuracy achieved by the CKKS FHE is 94.10% and the without implementation of CKKS FHE achieved the 96.60% with 10 number of communication rounds. Hence, it is proved that when applied the CKKS fully homomorphic encryption, it is a trade-off between performance and security.

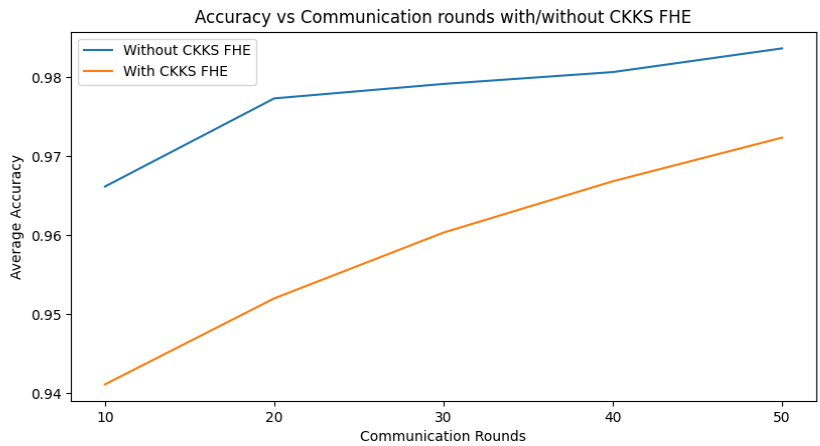

**Figure 9: Accuracy with and without CKKS FHE**

### 5.2.4 Cost of Operations

In Ethereum blockchain every transaction required the computational resources for execution and it costs some amount as gas fee according to the gas usage ratio. The cost of gas usage is dynamic per day and calculated in terms of minimum, average and maximum. In Figure 10, we have presented the gas usage ratio of eleven smart contract operation over Ethereum blockchain. The graph observation conforms that the gas usage is significantly high in some SC operations such as "add model hash" and "get rewards", whereas it is low for "task publisher registration", "client registration", "start forward bidding" and "selection of optimized clients". Furthermore, gas usage is moderate for some SC operations including "start auction" and "close auction". We did not compare the gas usage with random federated learning because there are no smart contract operations implemented in it. At the time of this experiment, we consider the gas cost of the operations at average and minimum rate such as 23.49 and 11.27, respectively. In Ethereum users can process transactions at their desired speed and cost is calculated accordingly. However, setting the value of gwei too low causes to stuck the transaction in blockchain due to insufficient gas fees. Hence, we calculated the gas price in average and minimum as transaction gas usage ratio.

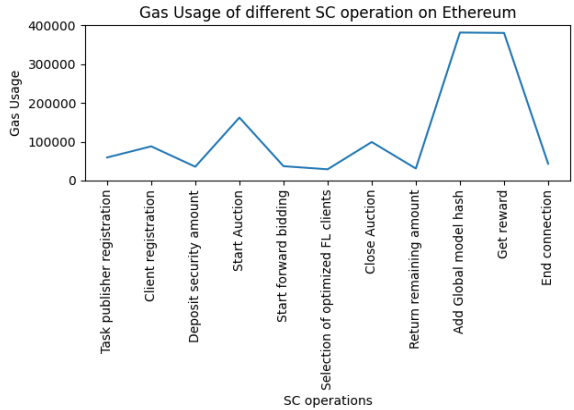
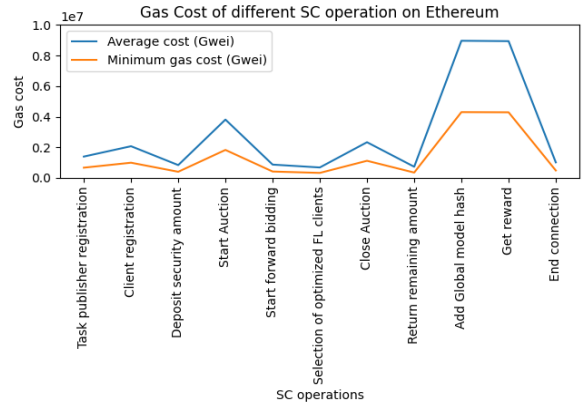

**Figure 10:** Gas usage by different SC operations over Ethereum blockchain network

**Figure 11:** Gas cost of different SC operations over Ethereum blockchain network

In Figure 11, we depicted gas cost at minimum and average rate according to gas usage ratio of different smart contract operations. Gas cost for "add model hash" and "get rewards" is highest, near to 8962515.54 calculated in gwei, used to upload the model to IPFS and retrieved the hash to get rewards. A gwei is the smallest unit of Ether and equals to the $10^9$ ETH. According to the complexity of SC operations, it required more transaction and execution fees. In our work, SC operations related to the "task publisher registration" and "client registration" have minimum gas cost, but it has great impact in terms of fair participation, without the registrations of clients or task publisher by SC cannot participate in model training.

### 5.2.4 Comparison with Baseline Studies

We compared our work with three baseline studies in terms of functionality comparison including fully homomorphic encryption, implementation of smart contracts, decentralization, and optimized selection of clients and penalty mechanism in Table 2.

**Baseline I [1]:** The work proposed by the McMahan et al. [1], with FedAvg scheme did not implement the additional privacy, smart contracts execution, and any penalty mechanism. Moreover, provide the model training with random selection of clients. However, we selected optimal clients to participate in FL training to achieve high accuracy.

**Baseline II [20]:** In the work of authors [20], proposed the FL-MAB can select the optimal clients by SC implementation and store the model at decentralized storage. However, did not provide the protection method for model such as homomorphic encryption.

**Baseline III [17]:** Authors [17], implemented the ElGamal homomorphic encryption which have only multiplicative homomorphic encryption, in FL we also required the additive HE to aggregate the local model updates into global model. No doubt a small modification is required to do for multiplicative into additive but cannot perform both operations at the same time. Apart from that, authors did not implement SC and optimized selection of clients.

**Baseline IV [36]:** In the work of authors [36], proposed a blockchain method for high quality model aggregation undertaken by the smart contracts. Besides, developed the method to evaluate the contribution of clients, however in case of privacy leakage attack did not provide the secure solution for local model updates.

**Baseline V [11]:** The work proposed by the Lo et al. [11], presented the blockchain based FL system to track and records the local model training by smart contract implementations. However, authors did not highlight the privacy mechanism and selected the clients for model training randomly.

**Our Paper:** We introduced the three kinds of smart contracts i.e., registration contract for clients and task publisher, forward bidding contract to select optimal client and payment settlement with rewards. Furthermore, we have implemented the fully homomorphic encryption i.e., CKKS to provide protection against privacy disclosure or if inference attack is happened then detected information is useless for adversary as it is encrypted.

Table 2: Functionality Comparison

| Functions/properties | Baseline I [1] | Baseline II [20] | Baseline III [17] | Baseline IV [36] | Baseline V [11] | Ours |
|---|---|---|---|---|---|---|
| FHE | No | No | Yes | No | No | Yes |
| Smart contract implementation | No | Yes | No | Yes | Yes | Yes |
| Decentralized storage | No | Yes | No | Yes | Yes | Yes |
| Optimized client selection | No | Yes | No | No | No | Yes |
| Penalty mechanism | No | No | No | No | No | Yes |

## 6. Conclusion and Future Work

In this paper, we have presented the blockchain based optimized client selection with privacy preserved manners for federated learning. A forward bidding smart contract is deployed to select the optimized FL clients for model training which replaced the traditional random selection of clients. Similarly, after FL model training, activated the payment settlement and reward smart contract for the engorgement of participated clients. Furthermore, we implemented the CKKS fully homomorphic encryption scheme to encrypt the local model updates before transmitting to the FL server. Compared to existing works, form experimental results, we achieved the better performance i.e., accuracy on benchmarked real-world datasets. With the integration of blockchain into federated learning provides a promising way to enable decentralization storage and transparent model history. In future work, we will consider the scenario of heterogeneous client's selection with some malicious clients, to separate them from normal clients and model training in decentralization federated learning system.

**CRediT authorship contribution statement**
Attia Qammar: Conceptualization, Methodology, Investigation, Writing-original draft & editing. Abdenacer Naouri: Software, Investigation, Visualization. Jianguo Ding: Review, Supervision, Project administration. Huansheng Ning: Resources, Supervision, Project administration

**Declaration of Competing Interest**
The authors declare that they have no known competing financial interests or personal relationships that could have appeared to influence the work reported in this paper.

**Data availability**
Public data is used.

## References

[1] B. McMahan, E. Moore, D. Ramage, S. Hampson, and B. A. y Arcas, "Communication-Efficient Learning of Deep Networks from Decentralized Data," in *Proceedings of the 20th International Conference on Artificial Intelligence and Statistics*, A. Singh and J. Zhu, Eds., in Proceedings of Machine Learning Research, vol. 54. PMLR, Jul. 2017, pp. 1273–1282. [Online]. Available: https://proceedings.mlr.press/v54/mcmahan17a.html

[2] T. Li, A. K. Sahu, A. Talwalkar, and V. Smith, "Federated Learning: Challenges, Methods, and Future Directions," *IEEE Signal Process Mag*, vol. 37, no. 3, pp. 50–60, May 2020, doi: 10.1109/MSP.2020.2975749.

[3] L. Lyu, H. Yu, J. Zhao, and Q. Yang, "Threats to Federated Learning," 2020, pp. 3–16. doi: 10.1007/978-3-030-63076-8_1.

[4] A. Qammar, J. Ding, and H. Ning, "Federated learning attack surface: taxonomy, cyber defences, challenges, and future directions," *Artif Intell Rev*, vol. 55, no. 5, pp. 3569–3606, Jun. 2022, doi: 10.1007/s10462-021-10098-w.

[5] Z. Wang, M. Song, Z. Zhang, Y. Song, Q. Wang, and H. Qi, "Beyond Inferring Class Representatives: User-Level Privacy Leakage From Federated Learning," in *IEEE INFOCOM 2019 - IEEE Conference on Computer Communications*, IEEE, Apr. 2019, pp. 2512–2520. doi: 10.1109/INFOCOM.2019.8737416.

[6] J. Kang, Z. Xiong, D. Niyato, H. Yu, Y.-C. Liang, and D. I. Kim, "Incentive Design for Efficient Federated Learning in Mobile Networks: A Contract Theory Approach," in *2019 IEEE VTS Asia Pacific Wireless Communications Symposium (APWCS)*, IEEE, Aug. 2019, pp. 1–5. doi: 10.1109/VTS-APWCS.2019.8851649.

[7] R. Zeng, S. Zhang, J. Wang, and X. Chu, "FMore: An Incentive Scheme of Multi-dimensional Auction for Federated Learning in MEC," in *2020 IEEE 40th International Conference on Distributed Computing Systems (ICDCS)*, IEEE, Nov. 2020, pp. 278–288. doi: 10.1109/ICDCS47774.2020.00094.

[8] T. Nishio and R. Yonetani, "Client Selection for Federated Learning with Heterogeneous Resources in Mobile Edge," in *ICC 2019 - 2019 IEEE International Conference on Communications (ICC)*, IEEE, May 2019, pp. 1–7. doi: 10.1109/ICC.2019.8761315.

[9] N. Bouacida and P. Mohapatra, "Vulnerabilities in Federated Learning," *IEEE Access*, vol. 9, pp. 63229–63249, 2021, doi: 10.1109/ACCESS.2021.3075203.

[10] Y. Li, C. Chen, N. Liu, H. Huang, Z. Zheng, and Q. Yan, "A Blockchain-Based Decentralized Federated Learning Framework with Committee Consensus," *IEEE Netw*, vol. 35, no. 1, pp. 234–241, Jan. 2021, doi: 10.1109/MNET.011.2000263.

[11] S. K. Lo *et al.*, "Toward Trustworthy AI: Blockchain-Based Architecture Design for Accountability and Fairness of Federated Learning Systems," *IEEE Internet Things J*, vol. 10, no. 4, pp. 3276–3284, Feb. 2023, doi: 10.1109/JIOT.2022.3144450.

[12] A. Qammar, A. Karim, H. Ning, and J. Ding, "Securing federated learning with blockchain: a systematic literature review," *Artif Intell Rev*, vol. 56, no. 5, pp. 3951–3985, May 2023, doi: 10.1007/s10462-022-10271-9.

[13] J. Zhu, J. Cao, D. Saxena, S. Jiang, and H. Ferradi, "Blockchain-empowered Federated Learning: Challenges, Solutions, and Future Directions," *ACM Comput Surv*, vol. 55, no. 11, pp. 1–31, Nov. 2023, doi: 10.1145/3570953.

[14] S. Singh, S. Rathore, O. Alfarraj, A. Tolba, and B. Yoon, "A framework for privacy-preservation of IoT healthcare data using Federated Learning and blockchain technology," *Future Generation Computer Systems*, vol. 129, pp. 380–388, Apr. 2022, doi: 10.1016/j.future.2021.11.028.

[15] M. Alloghani *et al.*, "A systematic review on the status and progress of homomorphic encryption technologies," *Journal of Information Security and Applications*, vol. 48, p. 102362, Oct. 2019, doi: 10.1016/j.jisa.2019.102362.

[16] F. O. Catak, I. Aydin, O. Elezaj, and S. Yildirim-Yayilgan, "Practical Implementation of Privacy Preserving Clustering Methods Using a Partially Homomorphic Encryption Algorithm," *Electronics (Basel)*, vol. 9, no. 2, p. 229, Jan. 2020, doi: 10.3390/electronics9020229.

[17] L. Zhang, J. Xu, P. Vijayakumar, P. K. Sharma, and U. Ghosh, "Homomorphic Encryption-based Privacy-preserving Federated Learning in IoT-enabled Healthcare System," *IEEE Trans Netw Sci Eng*, pp. 1–17, 2022, doi: 10.1109/TNSE.2022.3185327.

[18] J. K. Grewal, *ElGamal: Public-Key Cryptosystem*. Math and Computer Science Department, Indiana State University, 2015.

[19] nVotes, "Multiplicative vs Additive Homomorphic ElGamal," Jan. 01, 2020. https://nvotes.com/multiplicative-vs-additive-homomorphic-elgamal/ (accessed May 15, 2023).

[20] Z. Batool, K. Zhang, and M. Toews, "FL-MAB," in *Proceedings of the 37th ACM/SIGAPP Symposium on Applied Computing*, New York, NY, USA: ACM, Apr. 2022, pp. 299–307. doi: 10.1145/3477314.3507050.

[21] A. Fu, X. Zhang, N. Xiong, Y. Gao, H. Wang, and J. Zhang, "VFL: A Verifiable Federated Learning With Privacy-Preserving for Big Data in Industrial IoT," *IEEE Trans Industr Inform*, vol. 18, no. 5, pp. 3316–3326, May 2022, doi: 10.1109/TII.2020.3036166.

[22] Z. Su *et al.*, "Secure and Efficient Federated Learning for Smart Grid With Edge-Cloud Collaboration," *IEEE Trans Industr Inform*, vol. 18, no. 2, pp. 1333–1344, Feb. 2022, doi: 10.1109/TII.2021.3095506.

[23] M. Song *et al.*, "Analyzing User-Level Privacy Attack Against Federated Learning," *IEEE Journal on Selected Areas in Communications*, vol. 38, no. 10, pp. 2430–2444, Oct. 2020, doi: 10.1109/JSAC.2020.3000372.

[24] L. Zhu and S. Han, "Deep Leakage from Gradients," 2020, pp. 17–31. doi: 10.1007/978-3-030-63076-8_2.

[25] X. Luo, Y. Wu, X. Xiao, and B. C. Ooi, "Feature Inference Attack on Model Predictions in Vertical Federated Learning," in *2021 IEEE 37th International Conference on Data Engineering (ICDE)*, IEEE, Apr. 2021, pp. 181–192. doi: 10.1109/ICDE51399.2021.00023.

[26] L. Feng, Y. Zhao, S. Guo, X. Qiu, W. Li, and P. Yu, "BAFL: A Blockchain-Based Asynchronous Federated Learning Framework," *IEEE Transactions on Computers*, vol. 71, no. 5, pp. 1092–1103, May 2022, doi: 10.1109/TC.2021.3072033.

[27] H. Kim, J. Park, M. Bennis, and S.-L. Kim, "Blockchained On-Device Federated Learning," *IEEE Communications Letters*, vol. 24, no. 6, pp. 1279–1283, Jun. 2020, doi: 10.1109/LCOMM.2019.2921755.

[28] J. Kang, Z. Xiong, D. Niyato, S. Xie, and J. Zhang, "Incentive Mechanism for Reliable Federated Learning: A Joint Optimization Approach to Combining Reputation and Contract Theory," *IEEE Internet Things J*, vol. 6, no. 6, pp. 10700–10714, Dec. 2019, doi: 10.1109/JIOT.2019.2940820.

[29] J. Kang *et al.*, "Optimizing Task Assignment for Reliable Blockchain-Empowered Federated Edge Learning," *IEEE Trans Veh Technol*, vol. 70, no. 2, pp. 1910–1923, Feb. 2021, doi: 10.1109/TVT.2021.3055767.

[30] Y. Qu, S. R. Pokhrel, S. Garg, L. Gao, and Y. Xiang, "A Blockchained Federated Learning Framework for Cognitive Computing in Industry 4.0 Networks," *IEEE Trans Industr Inform*, vol. 17, no. 4, pp. 2964–2973, Apr. 2021, doi: 10.1109/TII.2020.3007817.

[31] B. Jia, X. Zhang, J. Liu, Y. Zhang, K. Huang, and Y. Liang, "Blockchain-Enabled Federated Learning Data Protection Aggregation Scheme With Differential Privacy and Homomorphic Encryption in IIoT," *IEEE Trans Industr Inform*, vol. 18, no. 6, pp. 4049–4058, Jun. 2022, doi: 10.1109/TII.2021.3085960.

[32] M. H. ur Rehman, A. M. Dirir, K. Salah, E. Damiani, and D. Svetinovic, "TrustFed: A Framework for Fair and Trustworthy Cross-Device Federated Learning in IIoT," *IEEE Trans Industr Inform*, vol. 17, no. 12, pp. 8485–8494, Dec. 2021, doi: 10.1109/TII.2021.3075706.

[33] Ethereum Foundation, "web3.py," 2023. https://web3py.readthedocs.io/en/stable/ (accessed May 15, 2023).

[34] Protocol Labs, "IPFS powers the Distributed Web," 2023. https://ipfs.tech/ (accessed May 16, 2023).

[35] J. H. Cheon, A. Kim, M. Kim, and Y. Song, "Homomorphic Encryption for Arithmetic of Approximate Numbers," 2017, pp. 409–437. doi: 10.1007/978-3-319-70694-8_15.

[36] J. Qi, F. Lin, Z. Chen, C. Tang, R. Jia, and M. Li, "High-Quality Model Aggregation for Blockchain-Based Federated Learning via Reputation-Motivated Task Participation," *IEEE Internet Things J*, vol. 9, no. 19, pp. 18378–18391, Oct. 2022, doi: 10.1109/JIOT.2022.3160425.